\documentstyle[epsfig]{article}

\newcounter{eqnletter}[equation]
\catcode`\@=11 \@addtoreset{equation}{section} \catcode`\@=12

\def\parno {\par\noindent}
\newcommand{\displayfrac}[2]{\frac{\displaystyle #1}{\displaystyle#2}}
\begin{document}
 \begin{centering}
{ \Large \bf Enumeration of simple random walks\\
and tridiagonal matrices.}\\
\vskip 1cm  G.M.Cicuta \\  Dept. of Physics, Univ. of Parma, and
INFN
\\ Viale delle Scienze,
 43100 Parma , Italy.
E-mail : cicuta@fis.unipr.it
\\ M.Contedini\\ Dept. of Physics, Univ. of Parma,
\\ Viale delle Scienze,
 43100 Parma , Italy.
E-mail : contedini@fis.unipr.it\\L.Molinari\\ Dept. of Physics, Univ. of
Milano and INFN, Sezione di Milano
\\ Via Celoria 16, 20133 Milano,
Italy. E-mail : luca.molinari@mi.infn.it

 \vskip 1cm
{\bf Abstract  }\\
 We present some old and  new results in the enumeration of random walks in
 one dimension, mostly developed in works of enumerative
 combinatorics. The relation between the trace of the $n$-th power
 of a tridiagonal matrix and  the enumeration of weighted paths of $n$
 steps allows an easier combinatorial enumeration
 of paths. It also seems promising for the theory of tridiagonal random matrices .
\end{centering}

\vskip 1cm

\section{Introduction. }
 Already at the foundation of the theory of random matrices by E.Wigner, 
the relevance of the combinatorics of random walks was recognized \cite{wig}.
The following decades witnessed an explosion of the theory and the 
applications of random matrices. A number of specific techniques were 
devised and the relation with the combinatorics of random walks was 
almost forgotten, with few 
remarkable exceptions \cite{sinai1}.
An interesting and recently investigated open problem, \cite{hn}, \cite{gk},
\cite{ze1}, \cite{ze2}, \cite{ze3}, \cite{der}, is the 
spectral density of non-hermitian tridiagonal random matrices. Here the 
enumeration of one dimensional random walks, where to each step 
an independent random variable is associated, plays a dominant role 
\cite{ccm}.  
Since the subject of random walks is very basic and useful, it has been 
thoroughly studied both by mathematicians and physicists for a long time, 
yet with techniques and aims so different that the relevant literature about 
inhomogeneous walks of one group ignores that of the other group.
   The aim of this paper is to review some old and recent results in 
enumerative combinatorics of random walks pertinent to the analysis of 
tridiagonal matrices in a language accessible to physicists, with examples, 
comments and relations that are not easily available in the literature. \\

A simple walk of $p$ steps on the lattice of integers, 
may be coded by a sequence $\{ \mu_1,.., \mu_p \}$, where each $\mu_k$ may
take the values $1$ (right move) or $-1$ (left move). The walk is
simple by the fact that we exclude the move $\mu_k=0$. The index $k$
can be regarded as discrete time. Since our walks start from the origin,
 at time $k$ the walker is on site $s(k)=\mu_1+..+ \mu_k$. It is very
convenient to view the walk as a continuous broken line in the
lattice $(k,s)$ of time and site occupancy. 
Most of the terminology is based on this
picture. An illustration of a simple walk we consider is given in Fig.1;  
the discrete time that numbers the steps is measured in the horizontal 
axis, and the  positions of the
 walker on the line are recorded in the vertical axis.

\epsfig{file=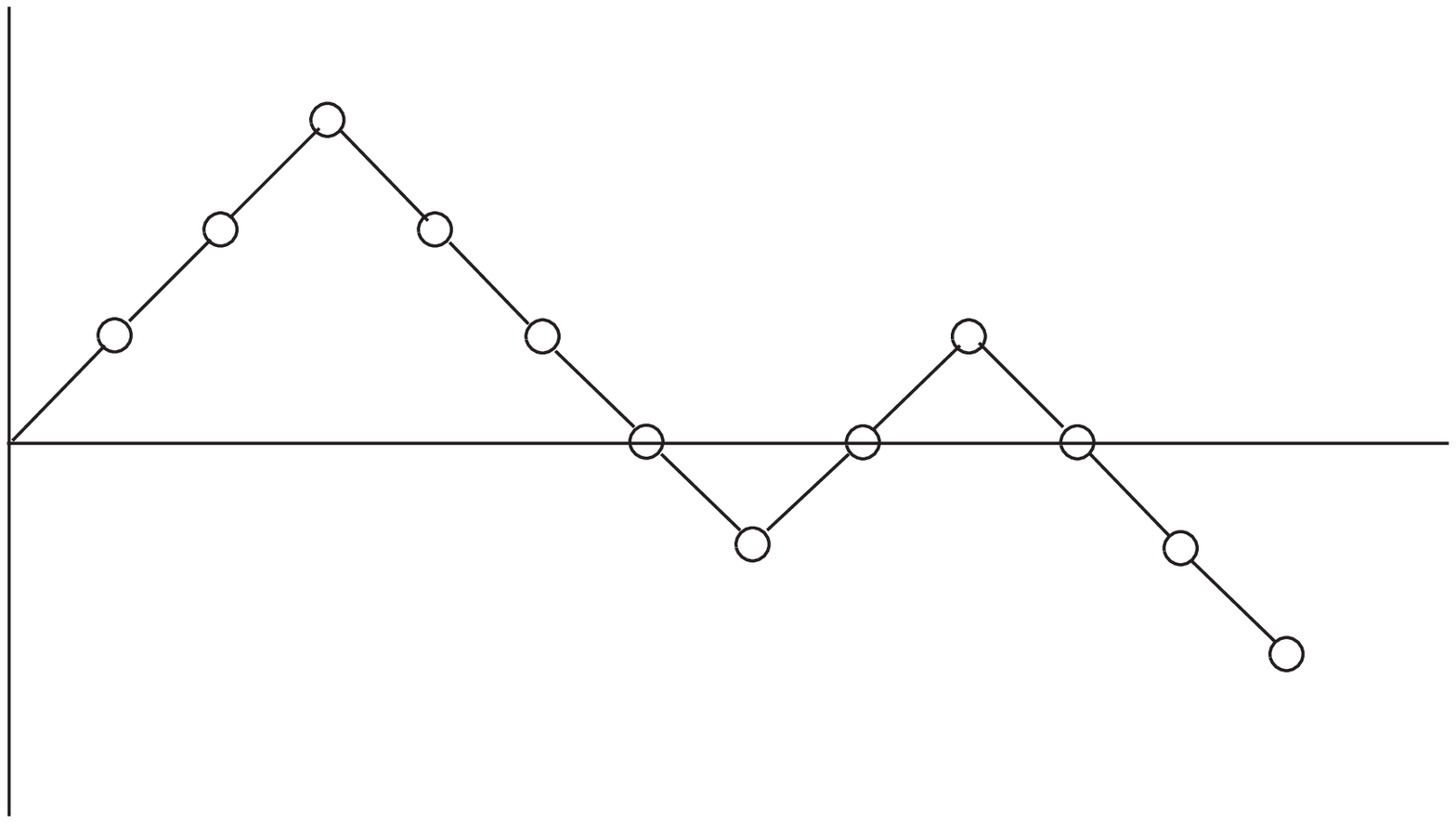,height=4cm}\\ Fig.1\\

A little terminology is useful. Simple walks where also the end site is
the origin, that is $s(p)=0$, are  "closed" . A simple closed walk is
 "weakly positive" if $s(k)\ge 0$ for all $k$. If we further require 
that it returns to the origin only
 at the end, the simple walk is "strictly positive".  Weakly positive
closed walks
are also known as Dyck paths.  The width of a walk is the number of different
visited sites.\\

Simple walks can be classified and then counted according to 
different properties.
A large number of results in the enumeration of simple walks have
been obtained by a method which we call  "iteration by length": 
the counting of walks with a given
property is related to the counting of similar but shorter ones, 
attached sequentially. In terms of
generating functions, the method leads to a system of equations
which, in several important cases, can be solved explicitly. One
of the best known examples is the enumeration of returns to the
origin in closed positive walks. In sect.2 we illustrate the
method by evaluating the number of peaks and valleys in several
ensembles of walks. It is a very simple exercise  given the known
methods in enumerative combinatorics \cite{deut} , \cite{goul} ,
\cite{stanley1}, and the method is familiar to physicists as it
recalls a Dyson equation in quantum field theory. \\

Recently we developed a technique , \cite{ccm}, to count the
number of visits of each site, for an ensemble of simple walks.
The method is again recursive but here properties of longer and
wider walks are computed in terms of shorter and narrower ones.
We  call this method "iteration by top insertions". It  leads to
explicit results in the form of sums of products of binomial
coefficients, the sum being actually in several indices taking 
values in the compositions of an integer, occasionally with further 
restrictions.
Since the results are involved, techniques to simplify them  would be very
useful. We think that this method is promising in the analysis of the
spectrum of tridiagonal random matrices and the related
random hopping problem in one dimension, where few exact results are
available, especially  in the non-hermitian case.
In this introduction we shall  describe in detail the connection of this
counting problem for simple walks with tridiagonal matrices.\parno
We then found that our method 'iteration by top insertions' 
had already been discovered in the form of enumeration
of vertices of planar rooted trees , \cite{goul} ,
\cite{stanley2}, \cite{mohanty}. We present this relation in sect.3 .

Occasionally a problem may be analyzed both by the simpler method
of sect.2 as well as by the more complex one of sect.3. We provide
an example by  evaluating  the area under Dyck paths, recently
discussed by Jonsson and Wheater \cite{jw} with the simple
method. While our discussion does not add to their solution, the
interpretation in terms of a tridiagonal matrix
facilitates its determination. \parno
The method in sect.2 is most useful 
for homogeneous (or almost homogeneous) random walks, where all steps 
(or almost all of them) have the same probability. Some early papers 
are \cite{rubin}, \cite{redner}. We remark that even the completely 
inhomogeneous walks where each step is associated with an independent 
random variable were enumerated \cite{raykin} by methods similar to 
the one we describe in sect.2. 
\\

As it is often the case,  computing the same quantities in two
different ways originates identities which could be hard to prove
in a direct way. For example, the sums over compositions of an
integer of products of binomials introduced in sect. 3, will be
shown to yield simple trigonometric sums. 
An early example of this type of sums is provided by eq.(22) 
in a paper by D.A.Klarner \cite{kla}.\\

Most of the methods described in this paper
have been developed for walks where also the move
$\mu_k=0$, that is an horizontal step , is allowed. Dyck paths
properly generalized to include such moves are called Motzkin
paths. Enumeration formulas become more complicated and, for the
introductory purpose of this paper, we avoid these walks.
Some results for generating functions for Dyck and Motzkin paths and their 
relation to continued fractions were recently summarized 
by C.Krattenthaler \cite{kra}.
\\

Ensembles of walks  are among the most basic
topics in probability theory and their statistical properties are
relevant in a large class of models in statistical mechanics.\\
In one dimension, to each  step $e_k$  of a walk $\gamma$  
 one may assign a weight $w(e_k)$. The whole path is then
attributed a weight  $w(\gamma)=w(e_1) w(e_2)..w(e_p)$. Finally
one is interested in summing over walks in some set $\Gamma $, 
weighted with their own factor $w(\gamma)$. One encounters this
procedure in the discrete approximation of functional integrals
$\int {\cal D}x \, f[x(t)]$, where the integration 
over continuous positive functions $x(t)$ with
boundary conditions $x(t_0)=0=x(t_f)$ is replaced by an 
ensemble of Dyck paths. The enumeration 
methods described here may be useful for the
evaluation of functional integrals where the paths may have
various restrictions. In the example 3 of sect.3, few cases of
weighted sums of Dyck paths are provided.\\

{\bf Random walks and tridiagonal matrices}\par
There is a very close relation between random walks and tridiagonal
matrices, and the latter are known to be a powerful tool for counting
walks with various specifications. Tridiagonal matrices are frequently
used to describe the motion of a particle in a one-dimensional chain.\parno
Since the walks we are considering are made of steps $\pm 1$,
we actually consider bidiagonal matrices of size $N\times N$ 
\begin{eqnarray}
M({\bf a},{\bf b}) = \pmatrix { 0 & b_1 & {}  & {} & {} & {} & {}
\cr
                                a_1 &  0  & b_2 & {} & {} & {}  & {} \cr
                                {} & a_2  &  0& \ldots & {} & {}  & {}\cr
                                {} & {}&  {}& {}     & {} & {} & {} \cr
                                {} & {}&  {}& {}     & {} & {} & b_{N-1}\cr
                                {} &{} &{}  & {}     & {} & a_{N-1} &  0  \cr }
\label{g.2}
\end{eqnarray}
\noindent

The starting point of our discussion  is the fact that the
explicit expansion of $M({\bf a}, {\bf b})^p_{i,j}$ immediately suggests to order
the various terms in
the sum according to some property of the random walk that connects site $i$
to site $j$ in $p$ steps.

In more detail, if we identify the matrix element $M({\bf a},{\bf b})_{ij}$
with the symbol $(i,j)$, and agree that repeated neighboring
indices are summed,
we have: $(M^p)_{ij}= (i,i_1)(i_1,i_2)\ldots (i_{p-1},j)$. Since the matrix
is bidiagonal, either $i_{k+1}=i_k -1$ or $i_{k+1}=i_k +1$. Because of this,
the pair $(i_k,i_{k+1})$ can also be identified with a move from site $i_k$ to
a neighboring site $i_{k+1}$ in a simple walk. Therefore, each  sequence
of $p$ factors $(i,i_1)\ldots (i_{p-1},j)$ not only represents a product of
matrix elements, but also a random walk of $p$ steps that connects 
site $i$ to site
$j$. The corresponding product of matrix elements may be interpreted as the
"weight" of the walk.

In this picture, the evaluation of $M({\bf a},{\bf b})^p_{ij}$ consists 
in  summing the weights of all  walks of $p$ steps from
$i$ to $j$. If we sum over lengths we get the resolvent,
which is the generating function for all weighted simple paths 
from $i$ to $j$:
\begin{eqnarray}
F(z,{\bf a}, {\bf b}; i,j) &=& \sum_p z^p M({\bf a},{\bf b})^p_{i,j}
         = [I-zM({\bf a},{\bf b})]^{-1}_{i,j}= \nonumber \\
         &=& (-1)^{i+j}\frac {\det [I-zM({\bf a},{\bf b}); j,i]
}{\det [I-zM({\bf a},{\bf b})]}
\label{g.55}
\end{eqnarray}
Here $[A ;i,j]$ is the matrix $A$ with row $i$ and column $j$
removed. This important result is very well known (see for
instance theor.4.7.2 in \cite{stanley1}).

It is very useful to note that the matrix  $M({\bf a},{\bf b})$ 
is similar to the matrix $M({\bf 1},{\bf x})$ where $x_i=a_i b_i$,
\begin{eqnarray}
M({\bf 1},{\bf x}) = \pmatrix { 0 & x_1 & {}  & {} & {} & {} & {} \cr
                                1 &  0  & x_2 & {} & {} & {}  & {} \cr
                                {} & 1  &  0& \ldots & {} & {}  & {}\cr
                                {} & {}&  {}& {}     & {} & {} & {} \cr
                                {} & {}&  {}& {}     & {} & {} &x_{N-1}\cr
                                {} &{} &{}  & {}     & {} & 1 &  0  \cr }
\label{g.3}
\end{eqnarray}
\noindent
In the relation $M({\bf a}, {\bf b})=S\, M({\bf
1},{\bf x}) \, S^{-1} $ the matrix $S$ is diagonal, with 
entries $s_1=1$, $s_2=a_1$, ... , $s_N = a_1a_2\ldots a_{N-1}$.
Therefore
\begin{eqnarray}
[M({\bf a},{\bf b})^p]_{ij} = \frac {s_i}{s_j} [M({\bf 1},{\bf x})^p]_{ij}
\label{g.5}
\end{eqnarray}

From this point, we exploit the similarity (\ref{g.5}), where we
note that the factor $s_i/s_j$ does not depend on $p$. To consider
the matrix $M({\bf 1},{\bf x})$, as we shall do hereafter, is a
great simplification. In each walk, only upward steps $(k,k+1)$
correspond to nontrivial factors $x_k$. Several walks may have
the same weight since they contain, but with different order, the
same intermediate steps $(k,k+1)$. By collecting these equal
contributions, we have a useful expression in terms of certain
counting numbers of random walks. The simplest example is
 
\begin{eqnarray}
[M({\bf 1},{\bf x})^{2p}]_{1,1} = \sum_{t=1}^{N-1}\sum_{\kappa (p,t)}
 N(n_1,\ldots ,n_t) x_1^{n_1}x_2^{n_2} \ldots x_t^{n_t} \label{l.56}
\end{eqnarray}
Here $N(n_1,\ldots ,n_t)$ is the number of Dyck paths that make
$n_j$ upward steps $(j-1,j)$ from site $j-1$ to site $j$, $j=1\ldots t$.
Their length is $2p=2(n_1+\ldots +n_t)$ and $t$ is the height, which is
bounded by the size of the matrix.\parno
The multiple sums in $n_1, \ldots, n_t\ge 1 $ such that
$n_1+\ldots +n_t=p$ are  summarized as the sum on the compositions of
$p$ into $t$ integers. We denote the set of compositions  of $p$ into
$t$ integers as $\kappa (p;t)$.\parno
This formula gives the explicit 
representation of $M({\bf 1}, {\bf x})^{2p}_{11}$ as a {\sl polynomial}
in the entries of the matrix.\parno
In the end of sect. 3 we describe the general 
case $[M({\bf 1},{\bf x})^{p}]_{i,j}$ which corresponds to the counting of 
random  paths of $p$ steps from site $i$ to site $j$, without positivity 
restriction. In the case  ${\bf x}={\bf 1}$ each walk has unit weight 
and one obtains the counting number of simple walks from site  $i$ to site 
$j$ of length $p$, restricted in the lattice
of points from $1$ to $N$, in terms of a simple matrix quantity.
Since eigenvalues and
eigenvectors are known, the multiple sums of the counting numbers
in eq.(\ref{g.6}) are equal to the trigonometric sum:
\begin{eqnarray}
[M({\bf 1},{\bf 1})^p]_{i,j}=\frac {2^{p+1}}{N+1} \sum_{k=1}^N
\cos^p \left ( \frac {k\pi}{N+1} \right ) \sin \left ( i\frac
{k\pi}{N+1} \right ) \sin \left (j \frac {k\pi}{N+1} \right )
\label{g.7}
\end{eqnarray}
 
One can use random bidiagonal matrices to count random walks with
restrictions. The trick is to assign  the variables $x_i$
a probability density such that the contribution of
unwanted walks is made to vanish in the average. \parno
For example, in \cite{ccm}  we studied the "q-root of unity" matrix
ensemble, generalizing a problem first suggested by Zee \cite{ze1}. 
A  random matrix $M({\bf 1},{\bf x})$ in the ensemble  
is characterized by independent random complex entries
$x_k$ chosen in the set of the $q-roots$ of unity,
with uniform probability. All moments $\langle x_k^r\rangle $ have
values 1 or 0, according to $r$ being a multiple of $q$ or not. 
Therefore, only walks that visit each site a number multiple of $q$ 
do contribute, and with weight one, to tr $M({\bf 1},{\bf x})^p$. 
This non local constraint strongly modifies the statistical properties
of the random walks, as it was shown Noh et al. \cite{noh} in a 
model of surface growth. 
The case $q=2$ of ``even-visiting walks'' has been investigated
in great detail by Bauer et al. \cite{bau}, while the
limit case $q=\infty$ was considered by Derrida et al. \cite{der}.
\par
On the other hand, given an ensemble of bidiagonal random matrices 
$M({\bf a}, {\bf b})$, the enumeration methods may be useful for the 
study of the spectral density, usually in the limit $N\to\infty$. 
If all random numbers $x_j=a_jb_j$  are positive, the
matrices are similar to real symmetric ones, $M({\bf c},{\bf c})$
with $c_j=\sqrt{x_j}$. The spectral density has support on the real axis and
can be achieved by the usual resolvent approach. The average resolvent
can be obtained by evaluating the ensemble average tr$M({\bf 1},{\bf x})^p$ 
for every $p$. To this end, the random walk description could  prove to be a 
valuable tool, even for approximate results, if one were able to acknowledge 
and count the dominant walks. In the case where the random variables 
$x_j$ are unrestricted, as in ref.\cite{ccm}, the spectral density has
complex support, and the evaluation of the resolvent may 
only allow the determination of the boundary of the support.\\

\section {Counting by iterations along the length of the path.}

One is often interested in the enumeration of paths according to a
parameter having the additive property: the parameter of a sequence
of Dyck paths is the sum, or a linear combination, of the parameters
of the single Dyck paths in the sequence, see for instance ref.\cite{deut}. 
Perhaps the best known example of additive parameter is the number of 
returns to the origin, see ref.\cite{kp1}. The method of enumeration by
iteration along the length is suitable for the case of additive
parameters, and it is here described following the pattern of
paper \cite{kp1}.\vskip 0.5truecm

{\bf  Statistics of peaks and valleys. }\\

Given  a walk we say that an
`inversion' occurs at time $k$ if $\mu_k \neq \mu_{k+1}$. One may
further partition inversions into peaks ($\mu_k=1$ ,
$\mu_{k+1}=-1$) and valleys ($\mu_k=-1$ , $\mu_{k+1}=1$ ). We
now show that the evaluation of the number of inversions in simple
random walks is a straightforward combinatorial exercise rather
similar to the evaluation of the number of visits to the origin,
which is known in the literature. The counting of inversions in 
long walks will be expressed in terms of the
counting in shorter and simpler ones.\\

It is useful to code a simple random walk of $n$ steps as a 
sequence of $m$ positive integers $n_1\ldots n_m$, that count
consecutive equal steps. A walk with initial step $\mu_1=1$
and described by the sequence $(n_1,\ldots ,n_m)$ begins with  $n_1$ 
steps $\mu=+1$, followed by $n_2$ steps in opposite direction $\mu=-1$,
and so on.  The sum $n_1+\ldots +n_m$ is the length $n$ of the walk,
and the number of inversions is $m-1$. For instance, the walk of $12$ 
steps and $3$ inversions  in Fig.1 is coded by the sequence $(3,4,2,3)$. 
The same sequence also describes a second walk, with opposite signs of 
steps.\\ 
Each sequence $(n_1, n_2,..,n_m)$ of positive integers is a
composition of the integer $n$ in $m$ parts and the number of such
compositions is $ {\cal C}(n,m)$
 \begin{eqnarray}
 {\cal C}(n,m)=  \left( n-1 \atop m-1 \right)
  \label{a.2}
\end{eqnarray}
It follows that the number of random paths of $n$ steps and $m$
inversions is $ 2 \,{\cal C}(n , m+1)$. By summing over inversions
one reproduces the total number of random paths with $n$ steps, 
$ 2 \sum_m {\cal C}(n,m)=   2^n $. \\

Our first evaluation is the number of walks of given length and
given number of inversions. It is convenient to consider first
walks returning to the origin (therefore the number of steps $n$
is even)  and introduce the counting numbers:\\
 - $c_1(n, m)$ is the number of strictly positive closed walks of 
$n$ steps and $m$ inversions,  $m=1, 3,..,n-3$\\ 
-  $c_2(n, m)$ is the number of weakly positive closed walks of $n$ 
steps and $m$ inversions, $m=1,3,..,n-1$ \\ 
-  $c_3(n, m)$ is the number of closed walks of $n$ steps
and $m$ inversions, with no positivity requirement, $m=1,
2,..,n-1$.\\

Since for strictly positive walks  it is $\mu_1=\mu_2=1$ and
$\mu_{n-1}=\mu_n=-1$, and for weakly positive walks 
it iss $\mu_1=1$ and $\mu_n=-1$, it follows that

 \begin{eqnarray}
c_1(n, m)&=& c_2(n-2, m)
 \quad , \quad {\rm for} \quad n \geq 2 \quad ;
 \nonumber \\
 c_1(2, m) &=& c_2 (2, m) = \delta_{m,1}
  \label{a.3}
\end{eqnarray}

\epsfig{file=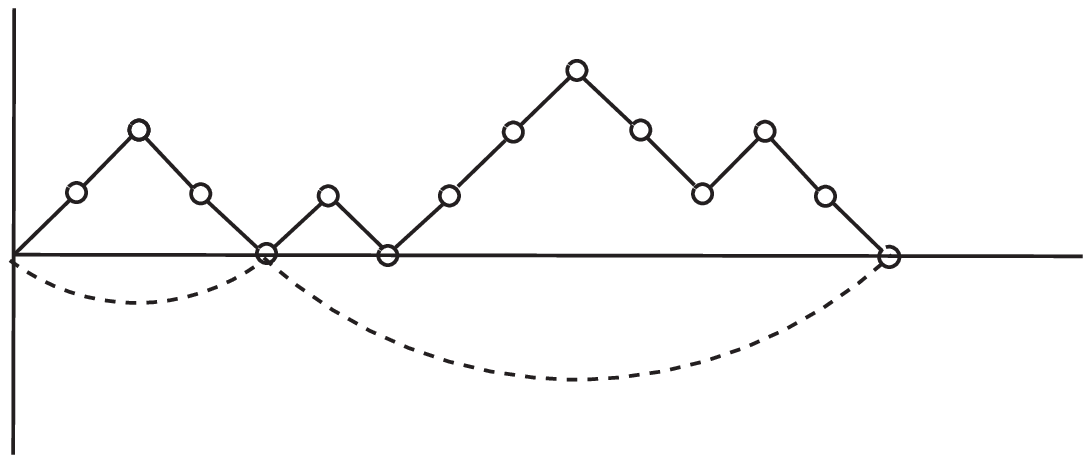,height=4cm}\\ Fig.2\\

A walk contributing to $c_2(n,m)$  either is in the class
contributing to $c_1(n, m)$ or has at least one visit to the
origin before the last site, (Fig.2).
If a walk contributing to $c_2$ but not to $c_1$ first returns to the
origin after $n_1$ steps, we have
 \begin{eqnarray}
c_2(n, m)=c_1(n, m)+\sum_{n_1=2,4,..,n-2 \atop m_1=3,5,..,n_1-3}
c_1(n_1, m_1) \, c_2(n-n_1, m-m_1 -1)
 \label{a.4}
\end{eqnarray}
In terms of generating functions
 \begin{eqnarray}
 C_i (x, y) = \sum_{n=2,4,..  \atop  m=1,2,..,n-1} c_i(n, m) \, x^n
 y^m
\label{a.5}
\end{eqnarray}
the eqs.(\ref{a.3}), (\ref{a.4}) translate into
 \begin{eqnarray}
 C_1(x, y)&=& x^2 y + x^2 C_2(x, y) \quad ; \nonumber\\
  C_2(x, y)&=& C_1(x,
 y)+y C_1(x, y)C_2(x, y)
\label{a.6}
\end{eqnarray}
The resulting algebraic quadratic equations lead to
\begin{eqnarray}
C_1(x, y)= \frac {1-x^2+x^2 y^2 - \sqrt{(1-x^2+x^2 y^2)^2 -4 x^2
y^2}}{2 y} \label{a.7}
\end{eqnarray}
\begin{eqnarray}
C_2(x, y)&=& \frac {1-x^2-x^2 y^2 - \sqrt{(1-x^2+x^2 y^2)^2 -4 x^2
y^2}}{2 x^2 y}
 \label{a.8}
\end{eqnarray}

The generating functions are closely related to the function $u(x, y)$ 
evaluated by Narayana
\cite{gessel}

\begin{eqnarray}
u(x, y) &=& 1-\sqrt{(1-x+xy)^2-4xy}= \nonumber\\ 
        &=& x+xy+ \sum_{n=2}^\infty
\sum_{r=1}^{n-1}  x^n y^r \frac{2}{n-1}
 \left( n-1 \atop r \right)
 \left( n-1 \atop n-r \right)
 \label{a.11}
\end{eqnarray}

For example, it is simple to obtain
\begin{eqnarray}
 C_2(x, y)=\sum_{n=2}^\infty \sum_{m=1}^{n-1}
 x^{2n-2}y^{2m-1}\frac{1}{n-1} \left(n-1 \atop m \right) \left(
 n-1 \atop n-m \right)
 \label{a.12}\end{eqnarray}

A walk contributing to $c_3(n,m)$ (an example is given in Fig.3)
either it is strictly positive or negative, or it has a first
return to the origin after $n_1$ steps, where it may or may not have an
inversion. \\

\epsfig{file=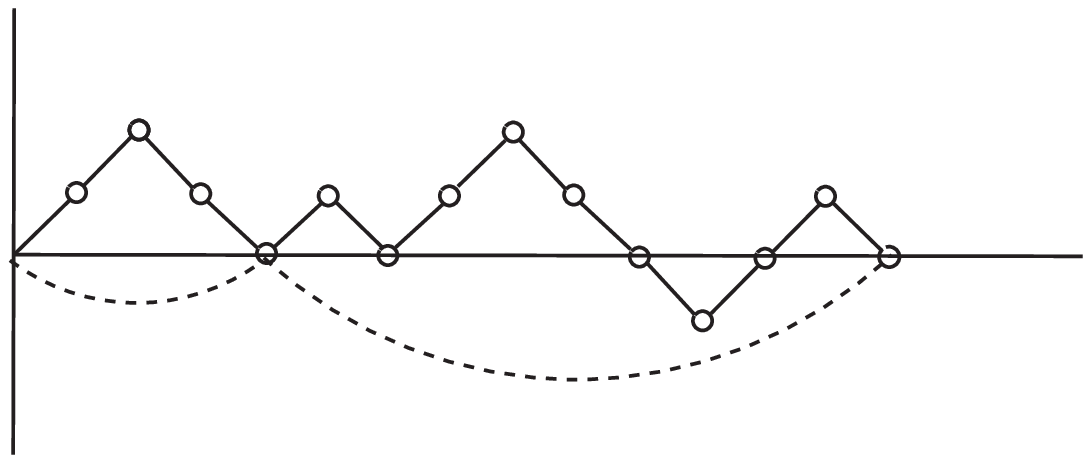,height=4cm}\\ Fig.3\\

This leads to the equation
\begin{eqnarray}
 C_3(x, y)= 2 C_1(x, y) + (1+y) C_1(x, y)  C_3(x, y)
 \label{a.9}
\end{eqnarray}
The solution is found using the previous results, and is written in the 
convenient form 
\begin{eqnarray}
 C_3(x, y) = -1 +\frac{1-x^2(1-y)^2}{\sqrt{(1-x^2+x^2 y^2)^2 -4 x^4
y^2 } } \label{a.10}
 \end{eqnarray}

To extract the coefficients of the power expansion we use the  
generating function of Legendre polynomials, also found in paper \cite{gessel}
\begin{eqnarray}
\frac{1}{\sqrt{(1-x^2-x^2y^2)^2-4x^4y^2}}&=& \sum_{n \geq 0}
x^{2n} (y^2-1)^n P_n \left(\frac{y^2+1}{y^2-1} \right) =
\nonumber\\ &=& \sum_{n,m} \left(n \atop m \right)^2 x^{2n} y^{2m}
\label{aa.2}
\end{eqnarray}

Then:

\begin{eqnarray}
 C_3(x, y)= 2 x^2y+2\sum_{n\geq 2,m \geq 1 }x^{2n} \left [
 y^{2m-1}
 \left( n-1 \atop m-1 \right)^2 +
 y^{2m}
 \left( n-1 \atop m \right)
 \left( n-1 \atop m-1 \right) \right ]\nonumber\\
\label{a.13}
\end{eqnarray}

Eq.(\ref{a.13}) has a simple interpretation when the
paths are coded by the sequences mentioned at the beginning of
this section. A random path with $n$ steps and $m$ inversions
contributing to $c_3(n,m)$ is coded by a sequence $(n_1, n_2,
..,n_{m+1})$ such that $\sum_{j \; odd} n_j= \sum_{j \; even} n_j
=n/2$. The two subsequences $s_{odd}=(n_1, n_3,..)$
and $s_{even}=(n_2, n_4,..)$ both have $(m+1)/2$ terms if $m$ is
odd, whereas if $m$ is even $s_{odd}$ has $\frac{m}{2}+1$ terms
and $s_{even}$ has $m/2$ terms. Then if $m$ is odd, the number of
sequences $(n_1, n_2, ..,n_{m+1})$ is given by ${\cal
C}(\frac{n}{2}, \frac{m}{2}+1)^2$,  whereas if
$m$ is even it is given by the product ${\cal C}(\frac{n}{2},
\frac{m}{2}+1) \; {\cal C}(\frac{n}{2}, \frac{m}{2})$. Since 
each sequence describes two walks, one obtains the coefficients 
$c_3(n,m)$ in eq.(\ref{a.13}).
Analogous results are quoted in the monograph \cite{sachov}.\\

Let us now consider random walks of $n$ steps and $m$ inversions without 
the restriction that the end point of the path is the origin, and no
restriction about positivity. Such walks correspond to all sequences
$(n_1, n_2,..,n_{m+1})$ with $n_1+\ldots n_{m+1}=n$. From eq.(\ref{a.2}) 
we obtain the generating function of the counting numbers:

\begin{eqnarray}
 C_4(x, y)= 2\sum_{n,m}  x^n \, y^m
 \left( n-1 \atop m \right)  =\frac{2}{(1+y)\, [1-x(1+y)]}
\label{a.14}
\end{eqnarray}

Finally we  evaluate the number $c_5(n,m)$  of weakly positive walks 
of $n$ steps and $m$ inversions, which may not be closed; an example is 
given in  Fig.4. The corresponding generating function is $C_5(x,y)$.

\epsfig{file=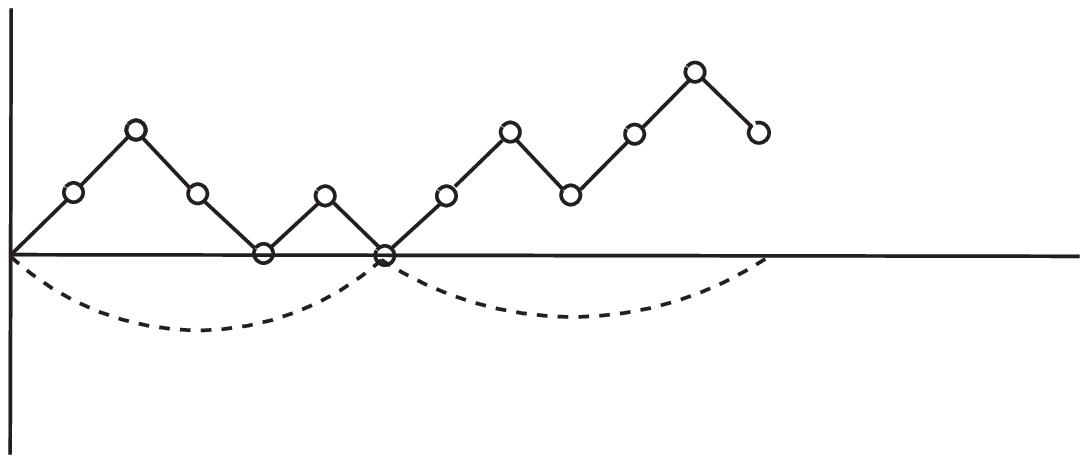,height=4cm}\\ Fig.4\\

\noindent
We may determine $c_5(n,m)$ in terms of shorter
walks, by partitioning the set of weakly positive walks in three disjoint
sets : a)  the set of walks that after an initial upward step 
never return to level one,
b) the set of walks with an initial upward step followed by a walk in
the ensemble described by the generating function $C_2(x,y)$; c) the 
set of walks of type b followed by a generic walk of type $C_5(x, y)$.  
This translates into the equation

\begin{eqnarray}
 C_5(x, y)= x+x\, C_5(x, y)+x\, C_2(x, y)+ x y \, C_2(x, y) \, C_5(x,y )
 \label{a.15}
\end{eqnarray}

Then
\begin{eqnarray}
 C_5(x, y)&=&  -\frac{1+x-xy}{2} +\frac{1}{2}
\frac{ \sqrt {(1-x^2+x^2y^2)^2-4x^2y^2} }{1-x-xy}  = \nonumber\\
&=& x y+\frac{1+x+x y}{2}\, C_3(x, y) 
\label{a.16}
\end{eqnarray}
The last equality allows an easy derivation of the counting numbers 
$c_5(n,m)$. As a simple check, we notice that the function 
$C_5(x,y=1)$, enumerates walks of fixed length and any number of inversions:
\begin{eqnarray}
 C_5(x, y=1)=  \frac{1}{2}\Bigg(-1 +\sqrt\frac{1+2x}{1-2x}
 \Bigg) \label{a.17}
\end{eqnarray}
\nonumber
The numbers are well known, see for instance \cite{kp1}.\vskip 0.5truecm

{\bf Statistics of the area.}\\

The  approach illustrated in the previous paragraph can also be used to
obtain  equations for the  generating functions that enumerate the positive
random walks of $T$ steps returning to the origin and enclosing with the
time axis a fixed area $A$. This problem was  recently discussed by
Jonsson and Wheater, and we refer to their paper \cite{jw}
for the physics motivations of the study and the analysis of the
functional equation for the generating function.\\

Let  $c_1(A, T)$ and $c_2(A,T)$ be the numbers of, respectively, strictly 
and weakly positive walks that return to the origin after $T$ steps, 
such that the area between the walk and the horizontal axis is $A$.
In both cases  $T$ is an even integer. Just as in Fig.2 and in 
eq.(\ref{a.4}) we have

 \begin{eqnarray}
c_2(A, T)=c_1(A, T)+\sum_{A_1=1,2,.. \atop T_1=2,4,..}
c_1(A_1, T_1) \, c_2(A-A_1, T-T_1 )
 \label{a.18}
\end{eqnarray}

Any weakly positive path may be elevated by adding a step at the beginning and
 another at the end, thus transforming it into a strictly positive path, then

\begin{eqnarray}
c_2 (A, T) &=& c_1(A+T+1, T+2) \qquad {\rm for} \quad A \geq 2 \quad ; \nonumber \\
c_2(A=1, T) &=& c_1(A=1, T)\; = \; \delta_{T ,2}
\label{a.19}
\end{eqnarray}

We define the generating functions
\begin{eqnarray}
F_i(x, y)=\sum_{A,T}
 c_i(A, T)  \;x^A  \, y^T  \qquad {\rm for} \quad i=1 \, , \, 2.
\label{a.20}
\end{eqnarray}
Then eqs.(\ref{a.18}), (\ref{a.19}) translate into
\begin{eqnarray}
F_2(x, y)&=& F_1(x,y)+F_1(x,y) \, F_2(x,y) \quad ; \nonumber \\
F_1(x, y)&=& x\,y^2+ x \, y^2 \,F_2(x, xy) \label{a.21}
\end{eqnarray}
One obtains the nonlinear functional equation of Jonsson and Wheater
\begin{eqnarray}
F_1(x,y)=x \,y^2+F_1(x,y)\,F_1(x, xy)
\label{a.22}
\end{eqnarray}
which has a formal solution as an infinite continued fraction
\begin{eqnarray}
F_1(x,y)= \displayfrac{x y^2}{1-\frac{x^3 y^2}{1-\displayfrac{x^5
y^2 }{1- \ldots  }}} \label{a.23}
\end{eqnarray}
In a similar way we obtain the functional equation for $F_2(x,y)$
\begin{eqnarray}
1-xy^2[1+F_2(x,y)]=  \frac{1}{1+F_2(x, xy)}   \label{a.24}
\end{eqnarray}
\begin{eqnarray}
F_2(x,y)= -1+1\displayfrac{1}{1-\frac{x\, y^2}{1-\displayfrac{x^3
\, y^2 }{1- \ldots  }}} \label{a.25}
\end{eqnarray}
From eqs. (\ref{a.22}), (\ref{a.24}), one  has $F_1(1,y)=\frac{1 -
\sqrt{1 - 4 y^2}}{2}$ and $F_2(1,y)=\frac{1-2 y^2-\sqrt{1-4
y^2}}{2 y^2}$. As shown in \cite{jw}, 
eq.(\ref{a.22}) allows the evaluation of $\frac{\partial}{\partial
x} F_1(x, y)|_{x=1}$, which is the generating function for the
area associated to the set of strictly positive paths of length
$T$
\begin{eqnarray}
\frac{\partial}{\partial x} F_1(x, y)|_{x=1}= \sum_{T} y^T 
\sum_A A\, c_1(A, T)   =\frac{y^2}{1-4 y^2}
 \label{a.26}
\end{eqnarray}
In the same way we obtain the generating function for the area
associated to the set of weakly positive paths of length $T$
\begin{eqnarray}
\frac{\partial}{\partial x} F_2(x, y)|_{x=1}= \sum_{T} y^T\, 
\sum_A A\, c_2(A, T)   =\frac{1-2y^2-\sqrt{1-4
y^2}}{2 y^2(1-4 y^2)}
 \label{a.27}
\end{eqnarray}
The results (\ref{a.26}), (\ref{a.27}) were obtained in
\cite{merlini} and extended to higher moments of the area in
\cite{chap} by a different method.  We do not proceed to the
analysis of eq.(\ref{a.22}), whose main
 properties were found in  paper \cite{jw}. In the next section
we describe a different approach to enumerate random walks,
which will provide a different viewpoint  of this problem.\\

\section {Counting by insertions which increase the width of the path.}

The counting method described in sect. 2 was based on
the construction of the ensemble of random walks by adding new
random walks at the end of previously considered ones. We now
present a different approach where the ensemble is generated by
inserting new random walks on the vertices of previously
considered  ones. \par
To evaluate the number of Dyck paths of $2n$ steps, height $t$, 
that visit prescribed numbers $N_j$ of times the sites $j=0,1\ldots t$, 
it is useful to consider the one-to-one map between Dyck paths and planar 
rooted trees \cite{bern}. The
correspondence is shown in Fig.5 where the rooted planar tree on the right 
side is mapped to the compressed "mountain range" of the left side,
where each  edge of the tree arises from one matched up-down
pair.\\ \noindent \epsfig{file=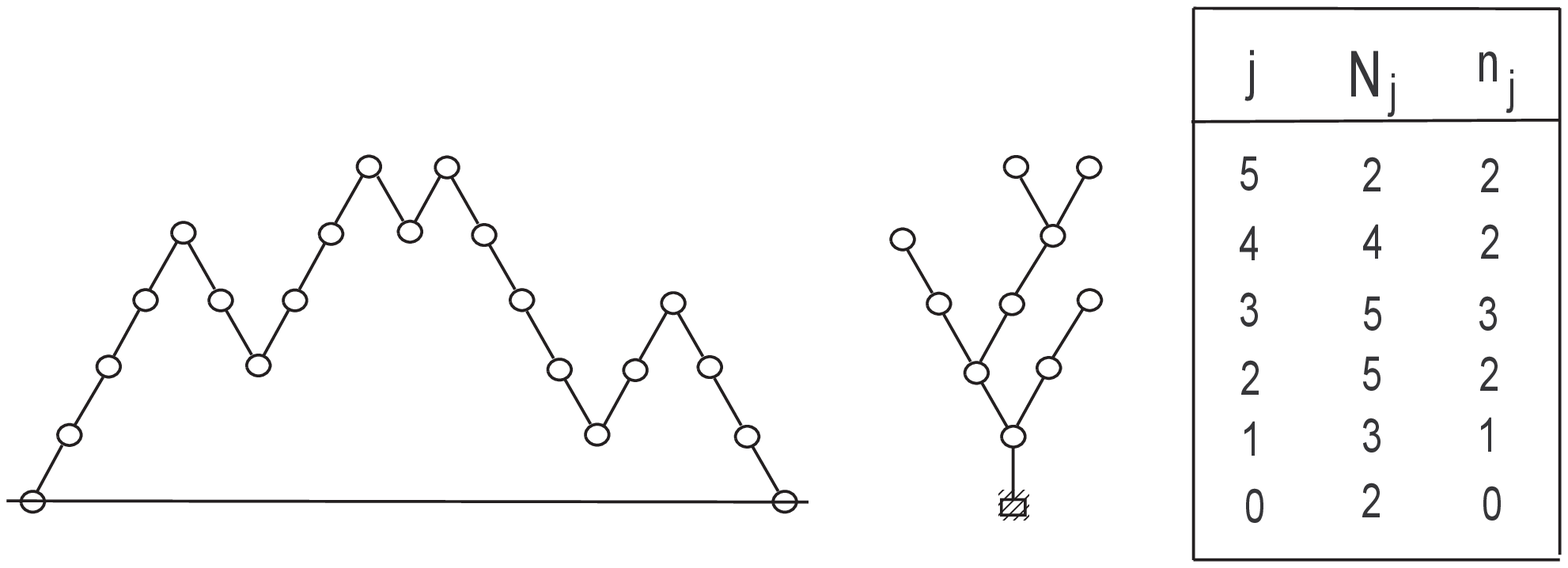,height=5cm}\\

\qquad \qquad \qquad  Fig.5\\

Given a Dyck path, let $n_j$ denote the number of up steps
between level $j-1$ and level $j$. Since $N_j$ is the sum of the
number of up steps from level $j-1$ and level $j$ plus the number
of down steps between level $j$ and level $j+1$, we have
\begin{eqnarray}
N_j=n_j+n_{j+1} \quad , \quad j=1,..,t \quad ; \quad n_0=0 \quad ,
\quad N_0=n_1+1 \label{r.1}
\end{eqnarray}

The following table recalls corresponding elements :\\

\noindent
\begin{tabular}{|l|l|} \hline
 Dyck path & planar rooted tree \\ \hline one pair of
matched up-down steps & one edge of the tree\\  length of path=$2n$
 & tree has $n$ edges\\ $N_j \quad , \quad j=1,.,t$ & sum
of degrees of vertices at level $j$\\ $ n_j \quad , \quad
j=1,.,t$ & number of edges between level $j-1$
 and level $j$ \\ $N_j-n_{j+1} $ \quad , \quad $j=1,..,t$ &$N_j-n_{j+1}= v_j=$
 number of vertices at level
 $j$ \\
  height $t$ &
height $t$\\ \hline
\end{tabular}

\vskip .5cm

The number $N(v_1, v_2,..,v_t)$  of planar rooted trees with given
set of non-root vertices $\{v_1,..,v_t\}$ is evaluated in a
recursive way because planar rooted trees with a set of non-root
vertices $\{v_1,..,v_t, v_{t+1}\}$ are obtained by adding
$v_{t+1}$ new vertices at level $t+1$ and using $t+1$ new edges to
connect all of them to some of the $v_t$ vertices at level $t$.
This may be done in $ \left( v_{t+1}+v_t-1 \atop v_{t+1} \right)$
ways. Then
\begin{eqnarray}
N(v_1, v_2,..,v_t)=\left( v_{2}+v_{1}-1 \atop v_{2}
\right)\left( v_{3}+v_{2}-1 \atop v_{3} \right)..\left(
v_{t}+v_{t-1}-1 \atop v_t \right) \label{r.3}
\end{eqnarray}

Since $v_j=N_j-n_{j+1}=n_j$ the evaluation (\ref{r.3}) provides
the number $N (n_1, n_2,..n_{t})$ of Dyck paths of height $t$,
with specified numbers $n_j$ of up steps between level $j-1$ and
level $j$
\begin{eqnarray}
N(n_1, n_2,..,n_{t})=\left( n_{1}+n_{2}-1 \atop n_{2}
\right)\left( n_{2}+n_{3}-1 \atop n_{3} \right)..\left(
n_{t-1}+n_{t}-1 \atop n_{t} \right) \label{r.4}
\end{eqnarray}
Such paths have total length  $2n$, where $n=n_1+..+n_t$.

 We remark that it might seem more natural to enumerate
Dyck paths of fixed length in terms of the integers $N_j$
enumerating the visits at each level, rather than in terms of the
positive integers $ n_j$ enumerating the  up steps from each
level. The second coding is superior because the positive numbers
$n_j $ are independent variables, save the restriction
$\sum n_j=n$.\\

We obtained the result (\ref{r.4}) by inserting new mountain ranges 
on top of the highest vertices of Dyck paths \cite{ccm}.  We were 
unaware of the previous works on the enumeration of vertices in 
planar rooted trees \cite{goul} , \cite{stanley2}. \\

We now proceed to discuss some properties of the enumeration numbers 
(\ref{r.4}) by introducing generating functions . They 
have a very direct relation with  general tridiagonal matrices. 
 This fruitful and interesting connection was described long ago
 by P.Flajolet, in his basic paper apparently almost ignored in later investigations
 \cite{Flajolet2}.\\ 

Let $F_t(x_1, x_2,..x_t)$ be the
generating function for the numbers $N(n_1,\ldots ,n_t)$:
\begin{eqnarray}
F_t(x_1, x_2,\ldots, x_t)=\sum_{n_j=1,.,\infty \; , \; j=1,.,t}N(n_1,
n_2,\ldots ,n_t)\,x_1^{n_1} x_2^{n_2}\ldots x_t^{n_t} \label{r.5}
\end{eqnarray}
We here derive various useful properties of the function, that
originate from a recursive relation.\\

{\underline {Proposition 1}}
\begin{eqnarray}
F_t(x_1,x_2,\ldots , x_t)= F_{t-1}(x_1, x_2,\ldots ,
\frac{x_{t-1}}{1-x_t}) -F_{t-1}(x_1, x_2,..,x_{t-1}) \label{r.6}
\end{eqnarray}
Proof:
\begin{eqnarray}
&&\sum_{n_j=1,.,\infty \; , \; j=1,.,t} N(n_1, n_2,\ldots ,n_t)x_1^{n_1}
\ldots x_n^{n_t}=\nonumber \\
&=&\sum_{n_j=1,.,\infty \; , \; j=1,.,t-1}N(n_1,
n_2,..,n_{t-1})x_1^{n_1} x_2^{n_2}..x_{t-1}^{n_{t-1}}
\sum_{n_t=1}^\infty \left( n_{t}+n_{t-1}-1 \atop n_{t} \right)
x_t^{n_t}= \nonumber\\
&=& \sum_{n_j=1,.,\infty \; , \;
j=1,.,t-1}N(n_1, n_2,..,n_{t-1})x_1^{n_1}
x_2^{n_2}..x_{t-1}^{n_{t-1}} \left[ (1-x_t)^{-n_{t-1}}-1 \right]=
\nonumber \\ &=& F_{t-1}(x_1, x_2,.., \frac{x_{t-1}}{1-x_t})
-F_{t-1}(x_1, x_2,..,x_{t-1})\quad \bullet \nonumber
\end{eqnarray}
\\

Beginning with $N(n)=1$, one iteratively constructs the first few generating
functions:
\begin{eqnarray}
F_1(x)=\frac {x}{1-x} \, ; \quad F_2(x,y)=\frac {xy}{(1-x)(1-x-y)}\, ;
 \nonumber
\end{eqnarray}
\begin{eqnarray}
 F_3(x,y,z)=\frac {xyz}{(1-x-y)(1-x-y-z+xz)} \, ;\quad \ldots \nonumber
\end{eqnarray}
They suggest the following formal solution to equation
(\ref{r.6}):\\

{\underline {Proposition 2}}
\begin{eqnarray}
F_t(x_1,x_2,...x_{t-1},x_t)\, &=&\, \frac
{x_1x_2..x_{t-1}x_t}{P_t(x_1,..,x_t) P_{t-1}(x_1,..,x_{t-1}) }= \nonumber \\
&=& \frac{P_{t-1}(x_2,..,x_t)}{P_t (x_1,..,x_t)} -\frac{P_{t-2}
(x_2,..,x_{t-1})}{P_{t-1}(x_1,..,x_{t-1})}\label{r.8}
\end{eqnarray}
where $P_t(x_1,..,x_t)$ is the polynomial generated through the recursion
\begin{eqnarray}
P_t(x_1,..,x_t) &=& P_{t-1}(x_1,..,x_{t-1}) - x_t \;
P_{t-2}(x_1,..,x_{t-2}) \nonumber\\ P_0=1 \; &,& \; P_1(x)=1-x
\label{r.9}
\end{eqnarray}
Proof: the first equality is proven by direct substitution of the
solution into the recursive formula (\ref{r.6}) and by using the
following identity, which is proven by repeated  use of eq.
(\ref{r.9}):
\begin{eqnarray}
P_t(x_1,..,x_t)=(1-x_t)P_{t-1}(x_1,..,x_{t-2}, \frac{x_{t-1}}{1-x_t}
) \label{r.10}\end{eqnarray}
To prove the second equality in eq.(\ref{r.8}), we formally 
solve the recurrence relation for
polynomials, with the given initial conditions, by means of a transfer
matrix:
\begin{eqnarray}
&&\pmatrix {P_t(x_1,..,x_t) & P_{t-1}(x_2,..,x_t)\cr
 P_{t-1}(x_1,..,x_{t-1})  & P_{t-2}(x_2,..,x_{t-1})\cr } =\nonumber\\
&& =\pmatrix {1 & -x_t\cr 1 & 0\cr }\pmatrix {1 & -x_{t-1}\cr 1 & 0\cr }
\ldots \pmatrix {1 & -x_1\cr 1 & 0\cr }\pmatrix {1 & 1\cr 1 & 0\cr }
\end{eqnarray}
By taking the determinant of both sides, we obtain a useful identity:
\begin{eqnarray}
 P_t(x_1,..,x_t) P_{t-2}(x_2,..,x_{t-1})-
 P_{t-1}(x_1,..,x_{t-1}) P_{t-1}(x_2,..,x_t) = -x_1x_2\ldots x_t
\nonumber\\
\label{r.99}
\end{eqnarray}
The equality follows by dividing both terms by $P_t(x_1,..,x_t)
P_{t-1}(x_1,..,x_{t-1})$. $\bullet$\\

Note that the recursive property implies that the polynomial
$P_t(x_1,..,x_t)$ coincides with the determinant of a matrix of size $t+1$:
\begin{eqnarray}
P_t(x_1,..,x_t)= \det [I+M({\bf 1},{\bf x})]
\label{r.100}
\end{eqnarray}

The generating function for walks with height not greater than $t$ is
\begin{eqnarray}
\Phi_t(x_1,..,x_t) &=& 1 + F_1(x_1) + F_2(x_1,x_2)+\ldots +
F_t(x_1,..,x_t)= \nonumber\\ &=& \frac
{P_{t-1}(x_2,..,x_t)}{P_t(x_1,..,x_t)}
 \label{r.11}
\end{eqnarray}
This function  has a simple representation
in terms of the bidiagonal matrix $M({\bf 1},{\bf x})$, of size $t+1$:
\begin{eqnarray}
\Phi_t(x_1,x_2,\ldots ,x_t) &=& [I+M({\bf 1},{\bf x})]^{-1}_{1,1}
\label{r.12}
\end{eqnarray}
By replacing the last term in the sum ({\ref{r.11}) with the right
side of eq. (\ref{r.6}) and resumming, one obtains the recursive
property
\begin{eqnarray}
\Phi_t(x_1,x_2,\ldots ,x_{t-1},x_t)=\Phi_{t-1}(x_1,x_2,\ldots,
\frac {x_{t-1}} {1-x_t}) \label{r.13}
\end{eqnarray}
and, since $\Phi_1(x)=(1-x)^{-1}$, the following finite
continued fraction representation:

\begin{eqnarray}
\Phi_t(x_1,x_2,\ldots ,x_t)=
\displayfrac{1}{1-\frac{x_1}{1-\displayfrac{x_2}{1- \ldots
\frac{x_{t-1}}{1-x_t} }}} \label{r.14}
\end{eqnarray}

The numbers $N(n_1,...,n_t)$ bring a very detailed information
about random walks. One is often interested in
reduced information, which require restricted summations over the
variables $n_1$...$n_t$. This is, in general, a rather formidable
task. We here provide some interesting examples. In the first we
count closed weakly positive walks of fixed height and length and
obtain identities involving sums of
products of binomials. In the second example, we again discuss the
problem of the area, by relating it to certain tridiagonal matrices. 
The third example addresses the issue of sums of weighted walks. 
\vskip 0.5truecm

{\bf {Example 1: enumeration of walks with fixed length and height.}}\parno
Let $C(n,t)$ be the number of weakly positive closed walks of length $2n$ and
height $t$. Evidently, the parameter $t$ cannot be larger than $n$. The sum 
over possible $t$ values gives the well known number of
weakly positive walks of length $2n$:
\begin{eqnarray}
\sum_{t=1}^n C(n,t)=\frac {1}{n+1} \left( 2n \atop n \right)  \label{t.1}
\end{eqnarray}
The generating function for the numbers $C(n,t)$ 
can be obtained  by setting  all arguments $x_i$ equal to $x$
in the function (\ref{r.5}):
\begin{eqnarray}
F_t(x)&=&\sum_{n=t,..,\infty} x^n C(n,t) =
\frac{x^t}{P_t(x) \, P_{t-1}(x)} =
\frac {P_{t-1}(x)}{P_t(x)}-\frac {P_{t-2}(x)}{P_{t-1}(x)}
\nonumber \\
C(n,t)&=&\sum_{\kappa (n;t)} \, N(n_1...n_t) ,\quad (n\ge t)
\label{t.2}
\end{eqnarray}
where, for shortness, we adopt the notation for the sum over the $t$
integers $n_i>1$ such that $n_1+...+n_t=n$ as a sum  over the set
$\kappa (n;t)$ of compositions of $n$ into $t$ integers $n_i\ge 1$.
 The polynomials $P_t(x)$ are easily evaluated from the recursion property
\begin{eqnarray}
  P_t(x)=P_{t-1}(x)- x P_{t-2}(x), \quad P_0=1, \,\, P_1(x)=1-x
\label{t.3}
\end{eqnarray}
 and have a simple expression in terms  of the roots  of the
equation $z^2-z+x=0$:
    \begin{eqnarray}
      P_t(x) &=& \frac {z_1^{t+2} - z_2^{t+2} }{z_1-z_2}= 
     \frac{1}{\sqrt {1-4x} }
\left [   \left (  \frac {1+\sqrt {1-4x} }{2} \right )^{t+2} -
\left (  \frac {1-\sqrt {1-4x} }{2} \right )^{t+2} \right ]
\nonumber\\ 
&=& \prod_{k=1}^{[(t+1)/2]} \left(1-4x \,\cos^2
\frac{k \pi}{t+2} \right) \label{t.4}
     \end{eqnarray}
For the last equality we used the known eigenvalues of the matrix
$M({\bf 1},{\bf 1})$. $P_t(x)$ is related to a Chebyshev polynomial 
of the second kind. \parno
The multiple sums over the compositions in eq.(\ref{t.2}) are obtained 
by extracting the coefficient of $x^n$ from the generating function $F_t(x)$. 
This is simple for small $t$; for example: 

  \begin{eqnarray}
 C(n,2)&=&\sum_{\kappa (n;2)} \left ( {{n_2+n_1-1} \atop {n_2}}
  \right )= 2^{n-1}-1 \quad ;\nonumber \\
  C(n,3)&=&\sum_{\kappa (n;3)} \left ( {{n_2+n_1-1}\atop  {n_2} }\right )
   \left ( {{n_3+n_2-1}\atop {n_3}}\right )= -2^{n-1}+ \nonumber \\
   &{}&+ \left(\frac{3+\sqrt{5}}{2}\right)^{n-2}\left(1+
 \frac{2}{\sqrt{5}}\right)+   \left(\frac{3-\sqrt{5}}{2}\right)^{n-2}
   \left(1-\frac{2}{\sqrt{5}}\right) \quad  ;
  \end{eqnarray}
We now consider the general case.\\

   {\underbar {Proposition 3}}\parno
    \begin{eqnarray}
 C(n,t) = -\frac{1}{t+2}\sum_j \frac {1-4a_j}{a_j^{n+1}} +
   \frac{1}{t+1}\sum_j \frac {1-4b_j}{b_j^{n+1}} \label{t.7a}
    \end{eqnarray}
     where $n\ge t$, $\{a_j\}$ and $\{b_j\}$ are, respectively, 
  the zeros of $P_t(x)$ and $P_{t-1}(x)$.\parno
    Proof: we use the following identity for the inverse of a polynomial 
$P(x)$ with simple  roots $x_j$:
    \begin{eqnarray}
   \frac {1}{P(x)} = \sum_{j=1}^n \frac {1}{P^\prime (x_j)}
   \frac {1}{x-x_j}   \nonumber
    \end{eqnarray}
    to obtain an expression for the inverse of the polynomial
 $P_t(x)P_{t-1}(x)$ in terms of the known roots  of the factors:
    \begin{eqnarray}
    \frac{1}{P_t(x)P_{t-1}(x)} =
\sum_j \frac {1}{P_t^\prime (a_j) P_{t-1}(a_j ) }\frac{1}{x-a_j }
    +\sum_j \frac {1}{P_t (b_j) P_{t-1}^\prime (b_j) }\frac{1}{x-b_j }
   \nonumber  \end{eqnarray}
 We then use the following results, which will be proven 
     \begin{eqnarray}
      P_t^\prime (a_j) P_{t-1}(a_j) = (t+2)\frac {a_j^t}{1-4a_j},
      \quad
  P_t(b_j) P_{t-1}^\prime (b_j) = - (t+1)\frac {b_j^t}{1-4b_j} \label{zzz}
      \end{eqnarray}
and obtain the power expansion of $F_t(x)$ we searched for:
     \begin{eqnarray}
     \frac{x^t}{P_t(x)P_{t-1}(x)} &=& \frac {1}{t+2}
     \sum_j \frac {1-4a_j}{a_j^t}
      \frac{x^t}{x-a_j}  - \frac{1}{t+1} \sum_j
   \frac {1-4b_j}{b_j^t}  \frac{x^t}{x-b_j}=\nonumber \\
  &=& \sum_{n=t}^\infty x^n  \left [
   -   \frac {1}{t+2}  \sum_j \frac {1-4a_j}{a_j^{n+1}} +
   \frac{1}{t+1} \sum_j\frac {1-4b_j}{b_j^{n+1}}
    \right ]  \nonumber
       \end{eqnarray}
 To derive the first identity in eq.(\ref{zzz}) one evaluates $P^\prime (x)$
in terms of the roots $z_1$ and $z_2$ of the equation $z^2-z+x=0$. For $x=a_j$
it is
      \begin{eqnarray}
        P_t^\prime (a_j) = -(t+2)\frac {z_1^{t+1}+z_2^{t+1} }
                      {(z_1-z_2)^2}
\nonumber
      \end{eqnarray}
The condition $P_t(a_j)=0$ implies that $z_1^{t+2}=z_2^{t+2}$; then:
  \begin{eqnarray}
 P^\prime_t(a_j)P_{t-1}(a_j) =  -(t+2)\frac {z_1^{2t+2}-z_2^{2t+2}
 }{[z_1-z_2]^3}= -(t+2)\frac {[z_1.z_2]^t
 \,[z_2^2-z_1^2] }{[z_1-z_2]^3}\nonumber
 \end{eqnarray}
This, together with $z_1+z_2=1$ and $z_1z_2=a_j$, allows the simplification 
that yields the result. The second identity is 
proven along the same line.$\bullet $\\

  The structure of the coefficients $C(n,t)$ as a two term difference
   suggests to introduce the  numbers  $C(n,\le t)$ of walks of length 
  $2n$ and height not larger than $t$:
  \begin{eqnarray}
  C(n,\le t) &=& C(n,1) + C(n,2)+\ldots C(n,t) =
  -\frac{1}{t+2}\sum_j \frac {1-4a_j}{a_j^{n+1}}= \nonumber\\
&=&
 \frac{2}{t+2} \sum_{j=1}^{t+1} \left( 4 \cos^2 \frac{j \pi}{t+2}
 \right)^{n} \, \sin^2 \frac{j \pi}{t+2}
  \label{t.7b}\\
&=&
2\sum_{\ell=- \left [ \frac {n}{t+2}\right ]  }^{\left [ \frac
{n}{t+2}\right ]}
 \left ( {2n}\atop  {n+\ell(t+2)} \right ) - \frac{1}{2}
 \sum_{\ell=-\left [ \frac {n+1}{t+2}\right ]    
}^{\left [ \frac {n+1}{t+2}\right ]}
 \left ( {2n+2}\atop {n+1+\ell(t+2)} \right ) \nonumber\\ \label{t.8}
  \end{eqnarray}
where  we inserted the explicit form for $a_j$ from
eq.(\ref{t.4}). The  form of  eq.(\ref{t.7b}) may be
recognized from the property $ C (n,\le t) =
 [M({\bf 1},{\bf 1})^{2n}]_{11} $, where $M({\bf 1},{\bf 1})$ has size
   $t+1$.  Of course $C(n,t) = C(n,\le t)-C(n,\le t-1)$.
 Note that $C(n,\le n)$ is precisely the number in eq. (\ref{t.1}).\parno
 The generating  function $\Phi_t (x)$ for the
 numbers is obtained  by setting
   all arguments equal to $x$ in the function (\ref{r.12}).
\begin{eqnarray}
\Phi_t(x)=\sum_{n=0}^\infty C(n,\leq t) \, x^n= \frac{2}{t+2}
\sum_{j=1}^{t+1} \frac{  \sin^2 \frac{j \pi}{t+2}
   }{1-4 x \cos^2 \frac{j \pi}{t+2} }
 \label{t.7c}
\end{eqnarray}
 \vskip 0.5truecm

{\bf {Example 2: enumeration of walks of fixed length and area}}.\parno
The area of closed weakly positive walks with numbers of upward steps
 $n_1,\ldots ,n_t$ is
$A=n_1+3n_2+5n_3 +\ldots +(2t-1) n_t$, while the total length is
$T=2n$, $n=n_1+\ldots +n_t$. With the position $x_1=xy$, $x_2=xy^3$,
..., $x_t=xy^{2t-1}$ in equation (\ref{r.5}) we obtain the
generating function $F_t(x,y)$ of counting numbers $C(n,t,A)$ of
positive closed walks with fixed length $2n$, height $t$ and area
$A$:
\begin{eqnarray}
F_t(x,y) &=& \sum_{n,A} x^n y^A \, C(n,t,A) \; = \nonumber\\
         &=&\frac{P_{t-1}(xy^3,..,xy^{2t-1})}{P_t(x y,..,xy^{2t-1})}
  - \frac{P_{t-2}(xy^3,..,xy^{2t-3})}{P_{t-1}(x y,..,xy^{2t-3})}
  \, ,\nonumber\\
 C(n,t,A)&=& \sum_{\kappa (n;t)} N(n_1,n_2,\ldots ,n_t)\,
 \delta (\sum_k (2k-1) n_k -A)
 \label{v.1}
\end{eqnarray}
where the $\delta $ function imposes the restriction that the integers
$n_1\ldots n_t$ of the compositions of $n$ should correspond to walks
with fixed area $A$. \parno
If we relax the restriction on height, being interested on counting numbers
$C(n,A)$ for the length and the area, the generating function is
\begin{eqnarray}
\Phi_\infty (x,y) &=& 1+\sum_{n,A} C(n,A) x^n y^A = 1+\sum_{t\ge 1}
 F_t(x,y)= \nonumber\\
  &=& \frac{P_\infty (xy^3,xy^5,\ldots )}{P_\infty (xy,xy^3,\ldots )}
\label{v.1b}
  \end{eqnarray}
 
We shall solve the counting problem by giving first the explicit expression of
the polynomial
  $P_t(x,y)=P_t (x y, xy^3, \ldots ,xy^{2t-1} )$, which coincides with the
determinant of the tridiagonal matrix of size $t+1$:
\begin{eqnarray}
I+M({\bf 1},x{\bf y})\,  = \,
 \pmatrix { 1 &  xy  &  {} &  {}            &  {}  &  {} & {} \cr
 1 &    1  &  xy^3          &  {}            &   {}  &  {} & {} \cr
 {} &    1  &  1               & xy^5        &   {}  &  {} & {}  \cr
 {} &   {}   &  \ldots        & \ldots      & \ldots & {} & {}  \cr
 {} &   {}   &    {} &    {} &  1      &  1  & xy^{2t-1} \cr
 {}  &  {}  &   {}            &  {}               & {}        &  1 & 1\cr }
\label{v.2}
\end{eqnarray}

{\underbar {Proposition 4}}\parno
\begin{eqnarray}
 P_t (x,y)= 1+\sum_{n=1}^{[(t+1)/2]} (-1)^n x^n y^{n(2n-1)}
 \prod_{k=1}^n \frac{1-y^{2(t-n+2-k)}}{1-y^{2k} } \label{v.3}
\end{eqnarray}
Proof: the polynomials solve the recurrence relation $P_t(x,y)=P_{t-1}(x,y) -
xy^{2t-1} P_{t-2}(x,y)$  with $P_0=1$ and $P_1(x,y)=1-xy$. The explicit
computation of the first few ones  suggests the following general structure:
$P_t(x,y)=\sum_{n=0}^{[(t+1)/2]} (-x)^n  c_{n,t}(y)$,
where $c_{n,t}(y)$ are polynomials for which recurrence relations are easily written.
By solving the first few ones, one easily arrives to conjecture the form stated
 in the proposition.
A formal proof then follows by induction, but we omit it.$\bullet $.

For the purpose of the statistics of the area, we need the limit $t\to\infty $:
\begin{eqnarray}
P_\infty (x,y) = 1 +\sum_{n>0} (-x)^n
\frac{y^{n(2n-1)}}{(1-y^2)(1-y^4)\ldots (1-y^{2n})} \label{v.4}
\end{eqnarray}
From this expression,  we derive the following functional equation:
\begin{eqnarray}
P_\infty (x,y) - P_\infty (xy^2,y) + xy \, P_\infty (xy^4,y) =0
\label{v.5}
\end{eqnarray}
Proof:
\begin{eqnarray}
P_\infty (xy^2,y)& = & 1 + \sum_{n>0} (-x)^n
\frac { y^{n(2n-1)}}{ (1-y^2)\ldots (1-y^{2n-2})}
 \frac {y^{2n}}{1-y^{2n}} =\nonumber\\
& = & P_\infty (x,y) -  \sum_{n>0} (-x)^n
 \frac {y^{n(2n-1)}}{(1-y^2)\ldots (1-y^{2(n-1) })} =\nonumber\\
&= & P_\infty (x,y) + xy \,P_\infty (xy^4,y) \qquad \bullet \nonumber
\end{eqnarray}
\noindent
Since $\Phi_\infty (x,y) = P_\infty (xy^2,y)/ P_\infty (x,y)$, 
we obtain a functional equation which corresponds to eq. (\ref{a.24})
for $F_2(x, y)$:
\begin{eqnarray}
\Phi_\infty (x,y) = 1+ xy \,\Phi_\infty (x,y) \, \Phi_\infty (xy^2,y)
\label{v.7}
\end{eqnarray}

The formula {\ref{v.1b} allows to compare $\Phi_\infty (x,y)$ with the
generating function $F_2(x,y)$  discussed in sect. 2, for weakly positive
walks. There, the
variable $y$ was conjugated to the total length $T=2n$ and $x$ was
conjugated to the area.
Here, for uniformity with the rest of the paper, $x$ is conjugated to
 $n$ and $y$ to $A$. It
turns out that $\Phi_\infty (x^2 ,y)=1+F_2(y,x)$.\\

Remark.  From the continued fraction representation of
$\Phi_{\infty} (x_1, x_2, x_3,...)$, see eq.(\ref{r.14})
\begin{eqnarray}
\Phi_{\infty}(x_1,x_2, x_3\ldots )=
\displayfrac{1}{1-\frac{x_1}{1-\displayfrac{x_2}{1-
\frac{x_3}{1-\ldots} }}} \label{z.1}
\end{eqnarray}
it is easy to see that it solves the formal equation
\begin{eqnarray}
\Phi_{\infty}(x_1,x_2, x_3,\ldots )=1+x_1 \,\Phi_{\infty}(x_1,x_2,
x_3,\ldots ) \, \Phi_{\infty}(x_2,x_3, x_4,\ldots )
 \label{z.2}
\end{eqnarray}
Let us suppose that each $x_k$ is a function of two variables
$x_k=x_k(x, y)$. The above formal equation will turn into a
functional equation for a function $\Phi_{\infty}(x, y)$ if
$x_k(x, y)=x_{k-1}(\overline{x},\overline{y})$, where 
the replacements $x\to \overline{x}$ and $y \to \overline{y}$ do not
depend on $k$. This is possible for the case
\begin{eqnarray}
x_k= x \, y^{p k+r} \quad , \quad p \; {\rm and} \; r \; {\rm
real} \label{z.3}
\end{eqnarray}
Then eq.(\ref{z.2}) becomes
\begin{eqnarray}
\Phi_{\infty}(x, y)=1+x y^{p+r} \, \Phi_{\infty}(x, y) \,
\Phi_{\infty}(x y^p, y) \label{z.4}
\end{eqnarray}
 Example 1 in this section corresponds to $p=0$ and $r=0$ ,
 Example 2 corresponds to $p=2$, $r=-1$. However the more general
equation (\ref{z.4}) does not add substantially to the previous
examples since it provides the generating function  for counting
Dyck paths with given number of steps and given $f(p, r)=\sum_j (p
j +r)n_j$, which is a linear combination of the area and the
length of the paths: $f(p,r)=(p/2)A+(2n)(r/2+p/4)$.\parno
Another interesting case is 
\begin{eqnarray}
 x_k=xy^{a^k}  \label{z41} 
\end{eqnarray} 
and will be discussed in the end of the next example.\parno
As this work was being written we became aware of recent papers
which have some overlap with this section \cite{fla} \cite{mans}
\cite{jani}.\\

{\bf Example 3 : Toy functional integrals.}\\ 
In the introduction we mentioned the subject of weighted Dyck paths
and speculated about the possibility of an operative definition of
path integral in the frame of Dyck paths. One may consider a variety 
of functionals $G[\gamma]$ where $\gamma$ is a Dyck path
of length $2n$, and write
\begin{eqnarray}
\int {\cal D} \gamma \, G[\gamma]=\frac{1}{C(n)} \sum_{\gamma \in
\Gamma(n)} G[\gamma] \nonumber
\end{eqnarray}
where $\{ s_k \}_{k=0}^{2n}$, with $s_0=s_{2n}=0$, is the sequence of
sites visited by the path $\gamma $ and  $C(n)$ is the cardinality of 
the set $\Gamma(n)$ of Dyck paths with $2n$ steps. It is given by
the Catalan number $C(n)=\frac{1}{n+1} \left( 2 n \atop n
\right)$ , see eq.(\ref{t.1}). Since we shall not 
consider the continuum limit, our finite sums are
here called toy functional integrals.\\
We are also interested in the set $\Gamma(n, t)$ of Dyck paths of
length $2 n $, height $t$  and cardinality $C(n,t)$, see Proposition 3,
whence one obtains the self-explanatory set $\Gamma (n,\le t)$.   
We discuss with some detail the class of functionals $G[\gamma]=\int d\tau 
g(\gamma (\tau ))$ and consider the toy integral 
\begin{eqnarray}
\int_{\Gamma(n,t)} {\cal D} \gamma \,\int d \tau
\,g(\gamma (\tau))&=&\frac{1}{C(n,t)} \sum_{\gamma
\in \Gamma(n, t)} \sum_{k=1}^{2n} g(s_k) \label{z.5}
\end{eqnarray}
Two interesting examples are
\begin{eqnarray}
 g(s) = s^r, \quad g(s)= c\, a^s, \quad a>0
\label{zz2} 
\end{eqnarray}
In particular, the toy integral with $g(s)=s$ evaluates the average 
area enclosed by Dyck paths of height $t$ and length $2n$. The second
example leads to explicit equations, and will be discussed after the
general case.\par
Eq.(\ref{z.5}) is first 
written in terms of the numbers $N_j$ of  visits of site $j$ of the 
Dyck path $\gamma$, next in terms of the familiar numbers of upward steps $n_j$
\begin{eqnarray}
\sum_{k=1}^{2n} g(s_k)=\sum_{j=0}^t g(j) N_j=\sum_{j=0}^t g(j)
(n_j+n_{j+1})=\sum_{j=0}^t[g(j)+g(j-1)] n_j \nonumber \\
 \label{z.7}
\end{eqnarray}
The sum over  paths is performed by introducing the counting
numbers $N(n_1, n_2,.., n_t)$
\begin{eqnarray}
 \sum_{\gamma \in \Gamma(n,t)} \sum_{j=1}^{t}
[g(j)+g(j-1)] n_j  \!\!&=&\!\! \sum_{\kappa (n;t)} N(n_1,
n_2,.., n_t)\sum_{j=1}^{t} [g(j)+g(j-1)] n_j= \nonumber \\ &=&
 \sum_{G \in {\cal G}} G \cdot C(n, t, G)
 \label{z.9}
\end{eqnarray}
where $C(n, t, G)$ is the number of Dyck paths of length $2n$,
height $t$ and fixed value $G$
\begin{eqnarray}
 C(n,t,G) =\sum_{\kappa (n;t)} N(n_1, n_2,.., n_t)\, \delta \bigg( G-\sum_i [
 g(i)+g(i-1)]n_i \bigg)
 \label{z.10}
\end{eqnarray}
and the sum $G \in {\cal G}$ runs over the set ${\cal G}$ of the
possible values of $G$. The generating function
$$ F_t(x,y)=\sum_{n, G} C(n,t,G)\,x^ny^G $$ 
is obtained from the general theory by setting $x_k =x\,y^{g(k)+g(k-1)}$ in
 eq.(\ref{r.8}). It may be evaluated from ratios of the
 determinants $P_t(x,y)$.\\
 The function 
\begin{eqnarray}
\Phi_\infty (x,y)=1+F_1(x,y)+
  \ldots+ F_t(x,y)+\ldots =\sum_{n, G}  C(n,G)x^n y^G
  \label{z.9b}
 \end{eqnarray}
 is the generating function for the numbers $C(n,G)=\sum_{t=1}^n C(n, t, G)$ 
that count Dyck paths of length $2n$ and fixed value $G$.
The generating function $f(x)$ for the numbers 
$G_n=\sum_{G \in {\cal G}} G  C(n,G) $ 
can be  written as the first derivative of $\Phi_\infty (x,y)$,
 \begin{eqnarray} 
 f(x)=\sum_{n=1}^\infty G_n x^n=\frac{\partial}{\partial y}
 \Phi_\infty (x,y) \Bigg|_{y=1}
  \label{z.9c}
 \end{eqnarray}
While we have been unable to find a "quadratic" functional equation for
the function $\Phi_\infty (x,y)$ related to the toy 
 integral (\ref{z.5}) with $g(s)=s^r$ for $r \geq 2$, we 
were successfull with the exponential function $g(s)=ca^s$. In this case
the functional evaluated on a Dyck path is, see eq. (\ref{z.7}),
   \begin{eqnarray}
 G = c \sum_{k=0}^{2n} a^{s_k} =
  c{{a+1}\over a}\sum_{s=1}^t a^s n_s
 \nonumber
 \end{eqnarray}
  The generating function for counting numbers is obtained with the
  position $x_k=xy^{(a^k)}$, so that
  $$
  x_1^{n_1}\ldots x_t^{n_t}=x^{n_1+\ldots n_t}y^{n_1a + n_2 a^2 +\ldots 
  +n_t a^t} = x^n y^G
  $$
  and, for the general theory, it is given by the continued fraction
 \begin{eqnarray}
  \Phi_{\infty}(x, y )=
\displayfrac{1}{1-\frac{x y^a}{1-\displayfrac{x y^{(a^2)}}{1-
\frac{x y^{(a^3)}}{1-\ldots} }}} \nonumber
\end{eqnarray}
which corresponds to the functional equation
\begin{eqnarray}
  \Phi_\infty (x,y) = 1+xy^a \Phi_\infty (x,y)\Phi_\infty (x,y^a)
 \label{z.13}
\end{eqnarray}
From this equation, since $ \Phi_\infty
(x,1)=(1-\sqrt{1-4x})/(2x)$, one easily obtains the generating
function for the numbers $G_n$, see eq.(\ref{z.9c}),
\begin{eqnarray}
 \frac{\partial}{\partial y}
\Phi_\infty (x,y)
\Bigg|_{y=1}=\frac{a}{x}\frac{(1-2x-\sqrt{1-4x})}{(1-a+(1+a)\sqrt{1-4x})}
 \nonumber
\end{eqnarray}
\\

It is interesting that the  counting numbers $C(n,A)$ , see eq.(\ref{v.1b}),
allow to define a partition function for Dyck paths of any length $L=2n$, 
weighted with their length $L$ and area $A$:
 \begin{eqnarray}
 \int_{\Gamma} {\cal D}\gamma \; e^{-a L[\gamma]-bA[\gamma]}=
   \sum_{L,A} e^{-a L -bA} C(L/2, A)=\Phi_\infty (e^{-2a},e^{-b})
  \label{z.11}
 \end{eqnarray}
 where $\Gamma $ is the set of all Dick paths and a different normalization
 has been used.\\

{\bf {Enumeration of closed paths.}}\\ 
 The counting numbers $N(n_1,\ldots ,n_t)$ apply to weakly positive
 closed paths, and are useful for expressing the matrix element (1,1) 
of $M({\bf 1},{\bf x})^{2p}$ as a polynomial in the entries $x_i$
of the matrix, see eq. (\ref{l.56}).\parno  
The generalization of the counting numbers to
arbitrary closed paths is straightforward, and is given in 
\\
\underbar {Proposition 5}\parno  
The number of closed paths of height $t$, depth $s$ and length $2p$,
containing $n_{-i}$ upward steps $(-i-1,-i)$, $i=s-1,\ldots ,0$, 
and $n_j$ upward steps $(j-1,j)$, $j=1,\ldots  ,t$, is
\begin{eqnarray}
N_{s,t} (2p;\, n_{-s+1}\ldots n_0 \,|\, n_1\ldots n_t)=\nonumber \\
=N(n_0,n_{-1}\ldots n_{-s+1})\left ( {n_0+n_1}\atop {n_1}\right )
N(n_1,n_2,\ldots ,n_t)=\nonumber \\
= \prod_{i=1}^{s-1} \left ( {n_{-i}+n_{-i+1}-1}\atop {n_{-i}} \right )
\left ( {n_0+n_1}\atop {n_0}\right ) \prod_{i=1}^t \left (
{n_i+n_{i+1}-1}\atop {n_{i+1}} \right )
\end{eqnarray}
Proof: Closed paths contributing to the above counting number are sequences 
of $n_1$ closed positive paths and $n_0$ closed negative paths. The factors
$N(n_0,\ldots ,n_{-s+1})$ and $N(n_1,\ldots ,n_t)$ respectively count 
the negative and positive closed paths that are obtained by joining the 
negative or positive subsequences. The intermediate binomial factor
counts the ways into which the subsequences can be ordered. $\bullet $
\\ 
The length $2p$ is specified in the symbol in view of a further
generalization to open paths; for closed paths 
$p=n_{-s+1}+\ldots +n_t$.\parno
The counting numbers allow to write the expression
for the matrix element $(i,i)$, $i=1\ldots N$ of the powers of the 
bidiagonal matrix $M({\bf 1},{\bf x})$ as a polynomial in the entries:
\begin{eqnarray}
M({\bf 1}, {\bf x})^{2p}_{ii} = \sum_{t=0}^{N-i}\sum_{s+t=1}^{N-1}
\sum_{\kappa (2p, s+t)} 
 N_{s,t}(2p;\, n_{-s+1}\ldots n_0\, |\, n_1\ldots
n_t)\nonumber \\
 x_{i-s}^{n_{-s+1}}\ldots x_{i-1}^{n_0} x_i^{n_1}\ldots 
x_{i+t-1}^{n_t}
\end{eqnarray}
\\

{\bf {Enumeration of general paths}}\parno
The  enumeration is extended to paths that may not return 
to the origin. For open paths the counting numbers 
$N_{s,t}(p\, ; \, n_{-s+1}, \ldots n_0 \, |\,
 n_1,\ldots n_t)$ correspond to an expression
which is more general than the above one.\parno
The paths have length $p$ and make the specified set of upward steps 
between sites $-s$ and $t$. Since 
the total number of upward steps is $u=n_{-s+1}+\ldots +n_t$ and the 
number of downward steps is $d=p-u$, the endpoint of the walks is site 
(level) $y=u-d=2u-p$. If $2u-p\neq 0$ the paths are open.\parno
For all paths it is $n_1,..,n_t\ge 1$, because the walk has to
reach the highest site $t$. However, for open paths some numbers $n_i$ with 
$i<0$ can be zero. More precisely, if $y\ge 0$ all $n_i\ge 1$ 
because the walk has to ascend from $-s$ to the level $y$; 
if $y<0$ it may be that $n_k=0$, for $k\le y$.\\

With this notation and postponing the problem of the
evaluation  of the counting numbers, we  write a general 
expression for the power of a matrix $M({\bf 1},{\bf x})$ as a
{\sl {polynomial in the entries}} $x_1,\ldots , x_{N-1}$:\\

{\underbar {Proposition 6}}
\begin{eqnarray}
[M({\bf 1},{\bf x})^p]_{i,j} = \sum_{s,t}
\sum_{n_{-s+1}\ldots n_t} N_{s,t}(p; \,
n_{-s+1}\ldots n_0 \, |   \,  n_1\ldots n_t) x_{i-s}^{n_{-s+1}}\ldots
x_{i-1}^{n_0} x_i^{n_1}\ldots x_{i+t-1}^{n_t} \nonumber \\
 \label{g.6}
\end{eqnarray}
In the first sum $s$ and $t$ have the following restrictions:
\parno
1)  the finite size of the matrix, which constrains walks
inside the strip of integers $1\ldots N-1$ that label the entries.
For this reason: $0\le s \le i-1$ and $0\le t\le N-1-i$.\parno
2) walks have to visit all sites between $i$ and $j$ included.
Therefore, if $j\ge i$ it is $t\ge j-i$; if $j\le i$ it is $ s\ge
i-j$. If $i=j$  both $s$ and $t$ cannot be zero.\parno
3) the sum on $n_{-s+1}\ldots n_t$ has the restriction that, being
$u=m_s+\ldots n_t$ the number of upward moves, the constraint
 $2u=j-i+p$ must be  satisfied. (Note
that for a bidiagonal matrix $[M^p]_{i,j}=0 $ if $p$ and $|j-i|$
have opposite parities).\\

The general expression for the counting numbers has been given
by Krattenthaler \cite{kra}, including the case of the null step.
We here discuss the case of paths that, starting from the origin, only 
visit sites $i>0$.\\

{\underbar {Proposition 7}}\parno
The number of positive walks of length $p$, that never return to
the origin and with prescribed numbers of upward steps is:
\begin{eqnarray}
N_{0,t}(p;\, 1, n_2,\ldots n_t)=\prod_{j=2}^y \left (
{n_j+n_{j-1}-2}\atop {n_j-1} \right ) N(n_y,\ldots , n_t)
\end{eqnarray}
where $y=2u-p$ is the height  of the
walk, $u=1+n_2+\ldots n_t$ is the total number of upward steps 
(see comment in proposition 5).\par   
To obtain this counting number it is convenient to generalize 
the planar rooted tree by using
edges of two types, with solid or dotted lines, as in Fig.6. There
is a one-to-one map between the "open mountain range" and the
corresponding generalized planar rooted tree: to any solid line
edge of the tree correspond, as before, a pair of matched
up-and-down sides of the mountain range, while a dotted line of
the tree corresponds to a up side of the mountain range without
the matching down side.

\noindent
\epsfig{file=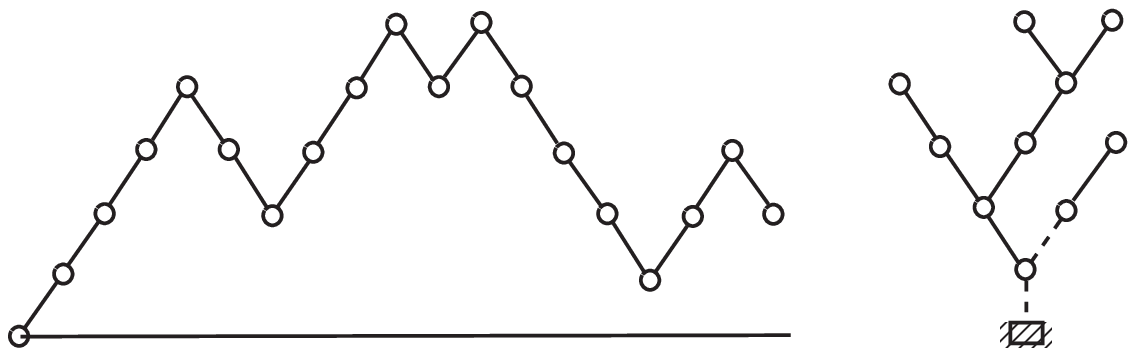,height=4.2cm}\\

\qquad \qquad \qquad  Fig.6\\

\noindent The correspondence between the number of visits at 
each level and the number of up steps is
\begin{eqnarray}
 N_j&=&n_j+n_{j+1}-1 \qquad , \qquad  1 \leq j \leq y \nonumber \\
 N_j&=&n_j+n_{j+1} \qquad \qquad, \qquad  y<j \leq t  \label{r.18}
\end{eqnarray}
Each vertex of the planar rooted tree is connected to a vertex at
 the lower level by a full line or a dotted line. $n_j$ counts
the full and dotted lines in the tree between level $j-1$ and level $j$; 
then
\begin{eqnarray}
 v_j=n_{j} \quad , \quad  j=1,..,t  \label{r.19}
\end{eqnarray}

The number of planar rooted trees is evaluated first by drawing
the root and a dotted path from the root to level $y$, with one
vertex at each level $1,..,y$, next by adding a number ${\tilde v_j}$
of new vertices at each level $j$, and connecting each vertex
with a full line to a vertex at the lower level. If we call
${\tilde v_j}=v_j$ for $j=y+1,..,t$ and ${\tilde v_j}=v_j-1$ for
$j=1,..,y$, we obtain that the number of planar rooted trees with
given set of non-root vertices:

\begin{eqnarray}
N(v_1, v_2,..,v_t)=\left( {\tilde v}_{t}+{ v}_{t-1}-1 \atop {\tilde v}_{t}
\right)\left( {\tilde v}_{t-1}+{ v}_{t-2}-1 \atop {\tilde v}_{t-1} \right)..
\left(
{\tilde v}_{2}+{ v}_1-1 \atop {\tilde v}_2 \right) \label{r.20}
\end{eqnarray}
From this, eq.(\ref{r.19}) and the relation between ${\tilde v}_j$
and $v_j$ we obtain the number in proposition 7. $\bullet $
\vskip 1cm

\begin{center}
{\bf Acknowledgments.}
\end{center}

We are very grateful to professors Sidney Redner, Christian 
Krattenthaler and  Emeric Deutsch  who provided 
us with comments, suggestions and relevant references after the
appearance of our preliminary draft, cond-mat/0011360, which 
eventually evolved into the present paper. \\


\begin{thebibliography}{99}

\bibitem{wig} E.Wigner, Characteristic Vectors of 
Bordered Matrices with Infinite 
Dimensions I, Ann. of Math. {\bf 62} (1955) 548-564.

\bibitem{sinai1} Ya.G.Sinai and A.B.Soshnikov, 
Central limit theorem for traces of 
large random symmetric matrices, Bol. Soc. Brazil. Mat., {\bf 29} (1998) 1-24;
A refinement of Wigner's Semicircle Law in a Neighborhood of 
the Spectrum Edge for  
Random Symmetric Matrices, Funct. Anal. and its Applic. 
{\bf 32} (1998) 114-131.

\bibitem{hn} N.Hatano, D.R.Nelson, Phys. Rev. Lett.{\bf 77} (1996) 570;
Phys. Rev. {\bf B56} (1997) 8651.

\bibitem{gk} I.Y.Goldsheid, B.A.Khoruzhenko, Distribution of eigenvalues
in non-Hermitian Anderson models,  Phys. Rev. Lett. {\bf 80} 
(1998) 2897 and Eigenvalue curves of asymmetric tridiagonal random
matrices, Electr. J. of Prob. {\bf 5} (2000) 1.

\bibitem{ze1} J.Feinberg, A.Zee, Non-hermitean localization and 
De-Localization, Phys.Rev. {\bf E 59} (1999) 6433-6443. 

\bibitem{ze2} E.Brezin, A.Zee, Non-hermitean delocalization: 
Multiple scattering and bounds,  
Nucl. Phys. {\bf B 509} (1998) 599-614.

\bibitem{ze3} C.Mudry et al., Density of states in the non-hermitian 
Lloyd model, 
Phys. Rev. {\bf B 58} (1998) 13539-13543.

\bibitem{der} B.Derrida, J.L.Jacobsen, R.Zeitak, 
Lyapounov exponent and density of 
states of a one-dimensional non-Hermitian Schrodinger equation, 
J. Stat. Phys. {\bf 98} (2000) 31-55.

\bibitem{noh} J.D.Noh, H.Park, M. den Nijs, Phys. Rev. Lett. {\bf 84}
(2000) 3891; J.D.Noh, H.Park, D.Kim, M. den Nijs, cond-mat/0103549;
D.S.Lee, M. den Nijs, cond-mat/0110485.

\bibitem{bau} M.Bauer, D.Bernard, J.M.Luck, Even-visiting random walks: exact
and asymptotic results in one dimension, J. Phys. A {\bf 34} (2001) 2659,
cond-mat/0102512.

\bibitem{ccm} G.M.Cicuta, M.Contedini, L.Molinari, Non-Hermitian tridiagonal
 random matrices and returns to the origin of a random walk, J. of Stat.Phys.
{\bf 98} (2000) 685-699.

\bibitem{deut} E.Deutsch,   Dyck path enumeration, Discrete Mathematics,
 {\bf 204} (1999) 167-202

\bibitem {goul} I.P.Goulden and D.M.Jackson, Combinatorial
Enumeration, John Wiley \& sons, 1983.

\bibitem {stanley1} R.P.Stanley, Enumerative Combinatorics,
Vol.1, Cambridge Univ. Press (1999).

\bibitem{rubin} R.J.Rubin, G.H.Weiss , 
Random walks on lattices. The problem of visit to 
a set of points revisited, J. Math. Phys. {\bf 23} (1982) 250-253.

\bibitem{redner} S.Redner, K.Kang, 
Asymptotic Solution of Interacting Walks in One 
Dimension, Phys. Rev. Lett. {\bf 51} (1983) 1729-1732; Unimolecular reaction 
Kinetics, Phys. Rev. A {\bf 30} (1984) 3362-3365.

\bibitem{raykin} M.Raykin, First-passage probability of a random 
walk on a disordered 
one-dimensional lattice, J. Phys. A {\bf 26} (1993) 449-466.

\bibitem{kla} D.A.Klarner, Correspondence between plane trees and binary 
sequences, 
J. Comb. Theory {\bf 9} (1970) 401-411.

\bibitem{kra} C.Krattenthaler, Permutations with restricted patterns 
and Dyck paths, 
math.Co/0002200.

\bibitem {bern} F.R.Bernhart, Catalan, Motzkin and Riordan numbers,
 Discrete Mathematics {\bf 204} (1999) 73-112.

\bibitem {stanley2} R.P.Stanley, Enumerative Combinatorics ,
Vol.2, Cambridge Univ. Press (1999)

\bibitem{mohanty} Sri G.Mohanty, Lattice Path Counting and 
Applications, Academic Press 1979.

\bibitem {jw} T.Jonsson, J.F.Wheater, Area Distribution for Directed
Random Walks, J. of Stat.Phys. {\bf 92} (1998) 713-725.

\bibitem {merlini} D.Merlini, R.Sprugnoli,M.C.Verri, The area
determined by under-diagonal lattice paths, Proc. CAAP, vol.1059,
Lecture Notes in Computer Science, (1996) 59-71.

\bibitem {chap} R.Chapman, Moments of Dyck paths, Discrete
Mathematics {\bf 204} (1999) 113-117.

\bibitem {kp1} P.Kirschenhofer, H.Prodinger, Return statistics of
simple random walks, Statistical Planning and Inference {\bf 54} (1996)
67-74.

\bibitem {gessel} I. Gessel, W. Goddard, W.Shur, H.S.Wilf, L.Yen,
Counting pairs of lattice paths by intersections, J. Comb. Theory,
A {\bf 74} (1996) 173-187, math/9409212.

\bibitem {sachov} V.Sachov, Probabilistic Methods in Combinatorial
Analysis, Encyclopedia of Mathematics and Its Applications 56,
Cambridge Univ. Press 1997, Chapter 3.

\bibitem{fla} P.Flajolet, F.Guillemin, The Formal Theory of
Birth-and Death Processes, Lattice Path Combinatorics and
Continued Fractions, (1999) INRIA report n.3667.

\bibitem{mans} T.Mansour, A.Vainshtein, Restricted permutations,
continued fractions and Chebyshev polynomials, (2000) The
Electronic Journal of Combinatorics {\bf 7} n.17, 1-9.

\bibitem{jani} M.Jani, R.G.Rieper, Continued fractions and Catalan
problems (2000) preprint math.CO/0001091.

\bibitem{Flajolet2} P. Flajolet, Combinatorial aspects of continued fractions,
Discrete Mathematics {\bf 32} (1980) 125-161.

\end{thebibliography}
\end{document}